\documentclass[10pt, conference]{IEEEtran}
\IEEEoverridecommandlockouts
\usepackage{cite}
\usepackage{amsmath,amssymb,amsfonts}
\usepackage{algorithmic}
\usepackage{graphicx}
\usepackage{textcomp}
\usepackage{xcolor}

\usepackage{hyperref}
\usepackage{pdfpages}
\usepackage{booktabs} 
\usepackage{placeins}
\usepackage{float}
\usepackage{afterpage}

\def\BibTeX{{\rm B\kern-.05em{\sc i\kern-.025em b}\kern-.08em
    T\kern-.1667em\lower.7ex\hbox{E}\kern-.125emX}}

\usepackage[a4paper, top=25mm, bottom=25mm, left=19mm, right=19mm]{geometry}

\begin{document}
\newgeometry{top=35mm, bottom=25mm, left=19mm, right=19mm}

\title{Beyond Speaker Identity: Text Guided Target Speech Extraction\\
\thanks{The work was carried out while the first author was an intern at Amazon Prime Video}
}

\author{
\IEEEauthorblockN{
Mingyue Huo\IEEEauthorrefmark{1}\IEEEauthorrefmark{2},
Abhinav Jain\IEEEauthorrefmark{2},
Cong Phuoc Huynh\IEEEauthorrefmark{2},
Fanjie Kong\IEEEauthorrefmark{2}
}
\IEEEauthorblockN{
Pichao Wang\IEEEauthorrefmark{2},
Zhu Liu\IEEEauthorrefmark{2},
Vimal Bhat\IEEEauthorrefmark{2}
}
\IEEEauthorblockA{
\IEEEauthorrefmark{1}\textit{University of Illinois Urbana-Champaign}, mhuo5@illinois.edu
}
\IEEEauthorblockA{
\IEEEauthorrefmark{2}\textit{Amazon Prime Video}, \{jaabhin, conghuyn, fanjikon, wpichao, zhuzliu, vimalb\}@amazon.com}
}


\maketitle

\begin{abstract}
 Target Speech Extraction (TSE) traditionally relies on explicit clues about the speaker's identity like enrollment audio, face images, or videos, which may not always be available. In this paper, we propose a text-guided TSE model StyleTSE that uses natural language descriptions of speaking style in addition to the audio clue to extract the desired speech from a given mixture. Our model integrates a speech separation network adapted from SepFormer with a bi-modality clue network that flexibly processes both audio and text clues. To train and evaluate our model, we introduce a new dataset TextrolMix with speech mixtures and natural language descriptions. Experimental results demonstrate that our method effectively separates speech based not only on \textit{who} is speaking, but also on \textit{how} they are speaking, enhancing TSE when traditional audio clues are absent. Demos are at: \url{https://mingyue66.github.io/TextrolMix/demo/}

\end{abstract}


\section{Introduction}
\label{sec:intro}

Target speech extraction (TSE) aims to isolate an individual's speech from a multi-talker mixture. Traditional methods rely on pre-recorded enrollment speech~\cite{Wang2019, vzmolikova2019speakerbeam, ge2020spex+}, but its unavailability and speaker variability pose challenges. Alternatives like synchronized speaker video, lip images, or microphone array data have been used to overcome these issues~\cite{ephrat2018looking, ochiai2019multimodal}. However, these clues often require pre-collection, limiting their practicality in real-world scenarios.

In contrast, text is easy to input and requires no prior collection, making it a practical and flexible solution.
Recently, advances in language models have popularized text-guided source separation, where natural language descriptions are used to extract a specific audio event or musical instrument from audio mixtures, yielding impressive results~\cite{kilgour2022text, dong2022clipsep, liu2023separate, li2023target, ma2024clapsep}. However, these models struggle with target ``speech" extraction due to two main drawbacks: (1) the dominance of non-speech audio event data in training, and (2) the difficulty of separating speech signals with similar spectro-temporal properties.
To the best of our knowledge, the first work applying natural language descriptions to TSE is LLM-TSE~\cite{hao2023typing}, which uses a large language model to extract semantic information from text queries for integration into a speech separation network. While LLM-TSE demonstrates the practicality of text-guided TSE, its performance is limited by the predefined range of attributes it explores, restricting broader applications.

To deal with the limitations of existing methods, we propose a flexible text-guided TSE model. To achieve this, we first recognize the necessity of a well-constructed speech-text paired dataset. Current approaches often rely on LLMs to create private corpora by rephrasing templates with predefined attributes, such as ``Extract voice with ⟨one specific characteristic⟩ from the mixture." However, these text clues are typically limited to a single attribute per instance, and focus more on rephrasing commands than on providing nuanced descriptions of speech characteristics~\cite{hao2023typing, jiang2024target}. Given the lack of publicly available datasets for text-guided TSE and the challenges posed by the open-ended and subjective nature of human perceptual voice description, we are motivated to develop a natural and diverse text-guided TSE dataset, TextrolMix. 

We also introduce StyleTSE, our text-guided target speech extraction model trained on the TextrolMix dataset. StyleTSE adapts the SepFormer~\cite{subakan2021attention, speechbrain} architecture with a bi-modality clue network that uses a gating mechanism to dynamically fuse text and audio clues, creating a more adaptive representation. Experimental results show that the proposed method is a robust and flexible solution for TSE, effectively handling scenarios with only audio clues, only text clues, or both.

\section{TSE Dataset: TextrolMix}
The proposed dataset TextrolMix consists of over 120,000 two-talker speech mixtures (157 hours), each with a pre-defined target speech, using audio clips from the TextrolSpeech dataset~\cite{ji2024textrolspeech}. To support a flexible text-guided TSE model, each mixture incorporates two types of clues: (1) a natural language description detailing the speaking style of the target speech, (2) a reference audio that shares specific style attributes with the target, accompanied by a corresponding text prompt. Examples are illustrated below.


Type I: ``Extract the male speaker with a shocked pitch and slow speaking rate."

Type II: A reference audio with surprised emotion, and ``Isolate the speech with the same emotion as the reference."



\begin{figure*}[h]
    \centering
    \includegraphics[page=1, width=\textwidth, trim=1cm 5cm 1cm 9cm, clip]{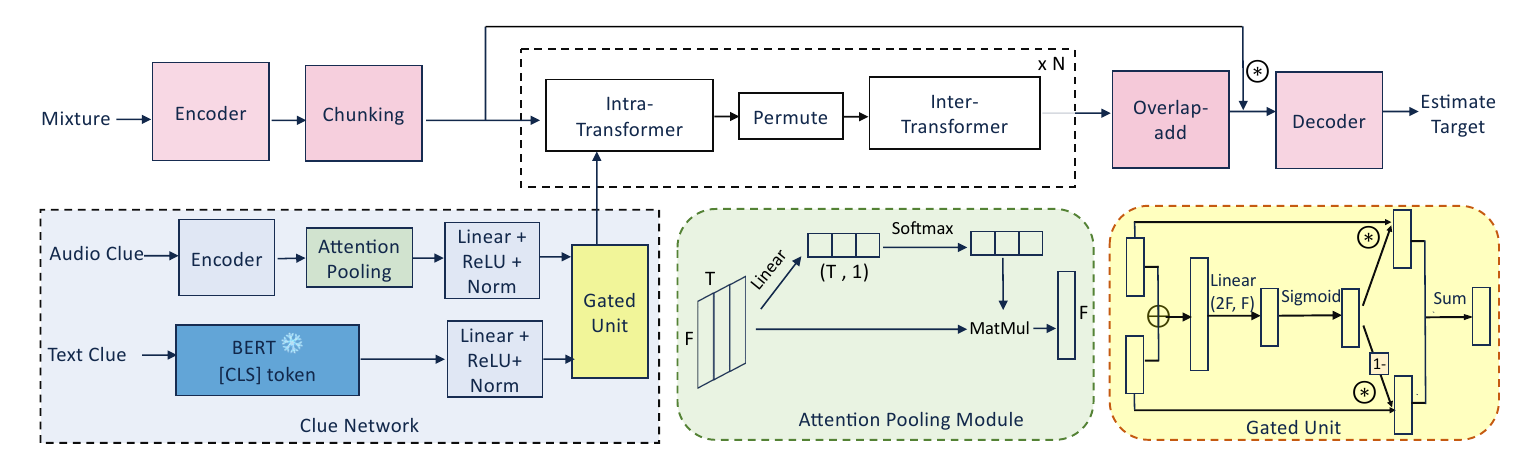}
    \caption{Proposed text-guided target speech extraction model StyleTSE features a separation network and a bi-modality clue network. The attention pooling module highlights specific time frames in the audio clue, while the gated unit dynamically combines text and audio clues. Symbols: $\circledast$ for element-wise multiplication, $\oplus$ for concatenation, and MatMul for matrix multiplication.}
    \label{fig:model}
\end{figure*}

The construction of TextrolMix involves two main steps: augmenting the original TextrolSpeech dataset and generating mixtures. TextrolSpeech was originally designed for text-to-speech tasks and included individual utterances with a single natural language description. We refined each utterance to include six attributes: speaker identity, emotion, pitch, gender, accent, and tempo. Furthermore, our preliminary experiments showed that models trained on longer text clues had poor generalizability to shorter, less descriptive texts. To counter it, we diversified the style descriptions into three lengths using templates. The longest provide comprehensive description like the example, whereas shorter versions highlight one or two attributes, such as ``The lady sounds happy" or ``High-pitched speaker." 


When creating mixtures, TextrolMix prioritizes speaking style over speaker identity. Unlike traditional TSE corpora such as WSJ-mix-extr~\cite{xu2019optimization}, we do not require mixtures to come from different speakers or use the target speaker for enrollment audio. Instead, each target utterance is paired with an interference utterance that differs in at least one style attribute such as emotion (e.g., a happy target paired with a sad interference). Additionally, the reference audio clue must share at least one attribute with the target but distinct from the non-targets. As a result, mixtures can include utterances from the same speaker if they exhibit distinguishable attributes, like different emotions. This design enables models trained on TextrolMix to extract target speech based on subtle attribute differences without requiring significantly distinct overall speaking styles.

Mixing adheres to the LibriMix procedure~\cite{cosentino2020librimix}. Speech utterances shorter than 3 seconds or longer than 15 seconds are excluded. The loudness of each utterance is uniformly adjusted between $-25$ and $-33$ dB Loudness Units Full Scale (LUFS), resulting in mixture Signal-to-Noise Ratios (SNRs) normally distributed with a mean of 0 dB and a standard deviation of $4.0$ dB. The TextrolMix dataset is split into training, development, and test sets at an 8:1:1 ratio. The dataset's metadata, generation scripts, and detailed statistics are accessible at \url{https://github.com/mingyue66/TextrolMix/}

\section{Text-guided TSE model}
\label{sec:model_strucutre}
Our proposed text-guided TSE model StyleTSE is developed based on a speech separation network, introduced in Sec.~\ref{ssec:separation_network}. A bi-modality clue network flexibly encodes both text and audio clues, and informs the separation network about the target speech, as detailed in Sec.~\ref{ssec:clue_network}. To effectively learn the complex voice characteristics, we incorporate dynamic mixing and a two-stage training process, explained in Sec.~\ref{ssec:train_strategy}.

\subsection{Separation network}
\label{ssec:separation_network}
Our separation network adapts SepFormer, a transformer-based model known for its state-of-the-art speech separation capabilities, to specifically extract the target utterance based on clues. This adaptation leverages SepFormer's robust dual-path framework for handling short- and long-term dependencies~\cite{luo2020dual}, enhancing target speech extraction.

As shown in Fig~\ref{fig:model}, the encoder-masker-decoder architecture begins with a 1-D convolutional encoder that transforms the time-domain mixture signal $x\in \mathbb{R}^L$ into an STFT-like representation $h\in \mathbb{R}^{F \times T}$. The encoded mixture is then chunked into $N_C$ overlapped time frames of size $C$, enhancing computational efficiency through parallel processing. Then, the masking network takes the chunked mixture representation $h' \in \mathbb{R}^{F \times C \times N_C}$ and clue embeddings, to be described later, to generate a mask that selectively filters the components of the target signal. The IntraTransformer and InterTransformer block repeats twice in our setup. For detailed architecture, see~\cite{subakan2021attention}. 
The masking network outputs chunked mask $h''\in \mathbb{R}^{F \times 1 \times C \times N_c} $, where $1$ indicates only one mask for the target is generated. After overlap-and-add and activation processes, the final mask $M \in \mathbb{R}^{F \times T}$ and the encoder output $h$ are multiplied element-wise. The decoder then uses a transposed convolution layer, mirroring the encoder’s stride and kernel size, to reconstruct the estimated target speech $\hat{s}$. 

\subsection{Clue network}
\label{ssec:clue_network}

The clue network is designed to encode the target speech's speaking style from clues, consisting of an audio clue encoder, a text clue encoder, and a gated fusion unit. 

The audio clue encoder has similar structure but independent weights with the mixture encoder. In contrast to pre-extracted x-vectors or d-vectors for speaker verification~\cite{hinton2012deep, snyder2018x}, the trainable encoder can capture not only speaker identity but also rich speaking style information, like emotion and accent. It transforms the time-domain audio clue into an STFT-like representation $A \in \mathbb{R}^{F \times T}$, which is then typically pooled to produce a time-invariant vector per utterance. Although average pooling, which treats all time frames equally, is common~\cite{vzmolikova2019speakerbeam, hao2023typing}, we found it inadequate for our task. Instead, we use an attention pooling module that dynamically calculates attention weights across time frames: $c_A = \sum_{t=1}^{T} w_t  A_{:,t}$, where attention weights $w \in \mathbb{R}^T$ is derived from a linear transformation and softmax, as highlighted in Fig.~\ref{fig:model}. The pooled audio clue vector undergoes a linear transformation, ReLU activation, and Layer Normalization to mitigate scale sensitivity, producing audio clue vector $c_A \in \mathbb{R}^{F'}$ that emphasizes speaking style from the reference audio clue. 

The text clue encoder employs a BERT-base transformer~\cite{devlin2018bert}. To preserve its robust pre-trained language understanding capabilities, we freeze the BERT model throughout the training process. It transforms text strings into a 768-dimensional [CLS] token, encapsulating condensed semantic information. Similarly, this embedding is linearly transformed, ReLU-activated, and Layer Normalized to produce a text clue vector $c_T \in \mathbb{R}^{F'}$ that matches the dimension and scale of the audio clue vector $c_A$. 

The audio and text clue vectors are first concatenated and then passed through a linear layer followed by a sigmoid function, which acts as a gating mechanism to generate dynamic weights. These weights determine how much information from each modality contributes to the final clue representation, resulting in a weighted sum. This gated fusion method has demonstrated superior performance in our experiments. Details on alternative fusion methods can be found in Sec.~\ref{subsec:ablation}.
In scenarios where either modality clue is unavailable---common in real-world applications---the model adapts accordingly during training and inference. When the audio clue is missing, a zero vector is used as a placeholder. For absent text clues, we employ an embedding from pseudo-text such as ``Extract the same speaker," treating audio clue as the identity enrollment as in the traditional TSE task.

Following the Exformer approach~\cite{wang2022semi}, the clue vector $c$ is expanded to match the dimensions of the chunked mixture representation $h'$ by replication and linear transformation, resulting in $C \in \mathbb{R}^{F \times C \times N_c}$. This embedding $C$ is then added to $h'$ at the start of each IntraTransformer,  allowing the separation network to condition target speech extraction on the clue embeddings.

\subsection{Training strategy}
\label{ssec:train_strategy}
The training objective is maximizing the scale-invariant signal-to-distortion ratio (SI-SDR)~\cite{vincent2006performance}, optimized to improve the accuracy of the estimated target $\hat{s}$ relative to the true target signal $s$. To improve the model’s robustness to various style attributes and adaptability to diverse clue inputs, we specifically employ dynamic mixing and a two-stage training process.

Dynamic mixing (DM) is a data augmentation technique that remixes utterances on-the-fly to generate more samples~\cite{subakan2021attention, wang2022semi, zeghidour2021wavesplit}. To enhance our model's ability to extract target speech using a wide range of style attributes beyond speaker identity, we apply DM in training. For the TextrolMix dataset, DM pairs utterances such that the interference shares the same style attributes as the predefined ones, preserving clue effectiveness. This approach yields an average of 2.45 alternative interference utterances per target. Additionally, utterances are mixed at random onset times to introduce further variability.

A two-stage training strategy is used to stabilize the clue network, inspired by~\cite{li2023target}. Our preliminary experiments revealed sub-optimal performance when training from scratch if either modality clue was missing, as the clue network struggled to produce stable audio and text embeddings. To address this, the first stage involves training with both audio and text clues—text specifically highlighting the speaking style from the reference audio. This initial phase, conducted around 100k steps at a higher learning rate without DM, focuses on developing and integrating the audio and text encoders. In the second stage, the model is trained with various clue inputs: text-audio, text-only, and audio-only in a 2:2:1 ratio with DM. This two-stage approach ensures gradual refinement of the clue encoders and robust performance across diverse input conditions.

\section{Experiments}
\label{sec:print}

\subsection{Training details}
We use the TextrolMix two-talker dataset sampled at 8 kHz for training and evaluation, truncating signals to 3 seconds and setting BERT's max token to 20. The separation network adopts SepFormer's optimal configurations. Both clue vectors are reduced to 256 dimensions before the gated unit. Training is performed on 8 NVIDIA V100 GPUs with a batch size of 56 for up to 200 epochs in total. We employ Adam optimizer, starting with a learning rate of 2e-4 for stage one, reduced to 1.5e-4 for stage two and halved if validation loss stalls for two consecutive epochs after 70 epochs, until it drops below 1e-6. 

\renewcommand{\arraystretch}{0.8}
\begin{table*}[ht]
\centering
\fontsize{9}{13}\selectfont
\caption{Model performances on the TextrolMix test set in SI-SDR improvement (dB) $\uparrow$ on the left and PESQ $\uparrow$ on the right. \#Param indicates trainable + frozen parameters in millions. Clue modality: Text (T), Audio (A); DM denotes dynamic mixing.}

\label{tab:model_comparison_subtasks}
\begin{tabular}{l l l  @{\hspace{1cm}} c @{\hspace{1cm}} c @{\hspace{1cm}} c @{\hspace{1cm}} c @{\hspace{1cm}} c}

\toprule

Model & Clue  &  \# Param &  Long  & Mid &   Short &  Avg &   \\
\midrule 
AudioSep \cite{liu2023separate} & T & 26.4 + 212  & 10.55 \quad 2.60 & 9.92 \quad 2.56 & 9.39 \quad 2.53 & 9.95  \quad 2.56 &  \\
LLM-TSE \cite{hao2023typing} & T & 5.2 + 110 & 10.67 \quad 2.58 & 10.76 \quad 2.56 & 11.34 \quad 2.59 & 10.92 \quad 2.58 & \\
StyleTSE  & T & 26.1 + 110 & 15.35 \quad 3.25 & 15.10 \quad 3.20 & 15.59 \quad 3.23 & 15.35 \quad 3.23 &  \\
StyleTSE + DM & T & 26.1 + 110 & \textbf{16.50 \quad 3.28} & \textbf{16.32 \quad 3.28} & \textbf{16.41 \quad 3.28} & \textbf{16.41 \quad 3.28} &  \\
\midrule

Model & Clue  & Speaker ID & Emotion & Accent & Pitch & Gender \\
\midrule
LLM-TSE \cite{hao2023typing} & A+T & 11.06 \quad 2.60 & 11.84 \quad 2.64 & 14.62 \quad 2.84 & 13.83 \quad 2.79 & 13.80 \quad 2.79 \\
StyleTSE  & A+T  & 14.36 \quad 3.27 & 14.81 \quad 3.29 & 19.04 \quad 3.53 & 18.30 \quad 3.48 & 17.73 \quad 3.45 \\
StyleTSE + DM & A+T & \textbf{15.76 \quad 3.29} & \textbf{15.95 \quad3.30} & \textbf{19.66 \quad 3.55} & \textbf{18.83 \quad 3.53} & \textbf{18.25 \quad 3.50} \\
\midrule

\end{tabular}
\end{table*}

\subsection{Results}

We evaluate the performance of StyleTSE using two key objective metrics: SI-SDR improvement (SI-SDRi) to measure the accuracy of speech separation, and Perceptual Evaluation of Speech Quality (PESQ)~\cite{rix2001perceptual} to assess the perceptual quality of the speech as it would be experienced by listeners.

\subsubsection{Text clue as style description} 
Table~\ref{tab:model_comparison_subtasks} upper section presents model performance when only a text clue describing the target speech's speaking style is provided. Introducing varied text lengths in the TextrolMix dataset resulted in consistent performance over 16 dB SI-SDRi, demonstrating our model's ability to interpret speaking style descriptions and accurately extract target speech.


For baseline comparison, we replicated the LLM-TSE model, which integrates text embeddings into a SpeakerBeam-like architecture that adapted from Conv-TasNet~\cite{vzmolikova2019speakerbeam, luo2019conv}. Since the original LLM-TSE code is unavailable, we used a frozen BERT model for the text encoder. Trained on the TextrolMix dataset using a two-stage strategy, our replication achieved performance close to the reported results---around 10 dB. 


We also fine-tuned the universal source separation model AudioSep~\cite{liu2023separate} with our proposed TextrolMix dataset. Starting from a 4 million step checkpoint, the model was continued trained for approximately 800k steps using their official scripts and individual speech utterances paired with varied length text descriptions. Initially, AudioSep's TSE performance was below 2 dB SI-SDRi, but fine-tuning significantly improved it, especially with longer text clues, highlighting the practical utility of the TextrolMix dataset.

\subsubsection{Text clue as attribute specifier}
Table~\ref{tab:model_comparison_subtasks} lower section shows our model's performance using both a reference audio clue and a corresponding text clue emphasizing specific style attributes. On average, our model achieves 16.84 dB SI-SDRi, performing better on accent, pitch, and gender, but weaker on speaker identity. In audio-only scenario, the audio clue is essentially treated as the enrollment for speaker identity as in the traditional TSE task. The performance was similar to the ``speaker identity" class when both audio and text clues are present ($\pm 0.05$ dB SI-SDRi), thus not reported separately.

These results suggest that using rich information from reference audio---traditionally considered ``harmful"~\cite{mu2024self}---can enable a more flexible TSE approach beyond speaker identity. The lower performance on speaker identity likely results from the dataset's greater voice variability caused by emotion, pitch, and tempo changes, which forces the model to prioritize these more distinctive cues.

\subsection{Ablation study on text-audio clue fusion}
\label{subsec:ablation}

We explored various fusion methods for combining audio and text clue embeddings, shown in Table~\ref{tab:ablation_study}. 
The average method assigns equal weights to the audio and text embeddings, summing them unless the audio is a zero vector, in which case only the text is used. The concat method simply concatenates both embeddings into a larger clue vector. The comparison among the first three rows highlights the effectiveness of the gated fusion method, which dynamically adjusts the contributions of audio and text clues. Furthermore, the last row shows a notable 2 dB drop when the attention pooling for processing audio clue is replaced with average pooling. This highlights the importance of dynamically weighing each frame to capture complex attributes from the audio clue, like emotion or accent, beyond just speaker identity.


\begin{table}[ht]
\centering
\caption{Ablation study on audio-text clue fusion method reported in SI-SDRi (dB)}
\label{tab:ablation_study}
\begin{tabular}{cccc}
\toprule
Fusion & Text-Audio &  Text-Only & Audio-Only \\
\midrule
Gated & {\textbf{16.84}} & {\textbf{16.41}} & {\textbf{15.72}} \\
Average & {15.89} & {15.48} & {14.78} \\ 
Concat & {15.84} & {15.46} & {14.61} \\
Gated w/o attpool & {14.86} & {14.95} & {13.73} \\

\bottomrule
\end{tabular}
\end{table}

\section{Conclusion}
We introduce a flexible text-guided target speech extraction model StyleTSE, and a multi-talker mixture dataset with natural language descriptions TextrolMix. StyleTSE demonstrates the ability to extract target speech by leveraging textual descriptions and differences in specific style attributes, even when the overall speaking style or speaker identity are not significantly distinct. Our two-stage training strategy effectively handles missing modalities, and dynamic mixing boosts performance across different input scenarios. Experimental results highlight the power of text as a guiding clue for target speech extraction, extending its scope beyond speaker identity to include a wider range of speaking styles.

\vfill\pagebreak

\bibliographystyle{IEEEtran}
\bibliography{refs}


\end{document}